\documentclass[a4paper]{eccomas_paper-2024}
\usepackage{graphicx}
\usepackage{hyperref}
\usepackage{graphicx}
\usepackage{amsmath}
\usepackage{amsfonts}
\usepackage{amssymb}

\usepackage[numbers]{natbib}

\title{OPTIMAL DESIGN OF VEHICLE DYNAMICS USING GRADIENT-BASED, MIXED-FIDELITY MULTIDISCIPLINARY OPTIMIZATION}

\author{HYUNMIN CHEONG$^1$, MEHRAN EBRAHIMI$^1$, HESAM SALEHIPOUR$^1$, ADRIAN BUTSCHER$^1$, AND ALEX TESSIER$^1$}

\heading{Hyunmin Cheong, Mehran Ebrahimi, Hesam Salehipour, Adrian Butscher, and Alex Tessier}

\address{$^{1}$ Autodesk Research\\
661 University Ave, Toronto, ON M5G 1M1, Canada\\
e-mail: \{firstname\}.\{lastname\}@autodesk.com}


\abstract{In automotive engineering, designing for optimal vehicle dynamics is challenging due to the complexities involved in analysing the behaviour of a multibody system. Typically, a simplified set of dynamics equations for only the key bodies of the vehicle such as the chassis and wheels are formulated while reducing their degrees of freedom. In contrast, one could employ high-fidelity multibody dynamics simulation and include more intricate details such as the individual suspension components while considering full degrees of freedom for all bodies; however, this is more computationally demanding. Also, for gradient-based design optimization, computing adjoints for different objective functions can be more challenging for the latter approach, and often not feasible if an existing multibody dynamics solver is used.

We propose a mixed-fidelity multidisciplinary approach, in which a simplified set of dynamics equations are used to model the whole vehicle while incorporating a high-fidelity multibody suspension module as an additional coupled discipline. We then employ MAUD (modular analysis and unified derivatives) to combine analytical derivatives based on the dynamics equations and finite differences obtained using an existing multibody solver. Also, we use a collocation method for time integration, which solves for both the system trajectory and optimal design variables simultaneously. The benefits of our approach are shown in an experiment conducted to find optimal vehicle parameters that optimize ride comfort and driving performance considering vertical vehicle dynamics.}

\begin{document}
\thispagestyle{empty}

\section{INTRODUCTION}

Optimal design of complex systems, such as aircraft, automobiles, and buildings, requires considering multiple subsystems that interact with each other. To solve such problems, multidisciplinary design optimization (MDO) methods have been developed where the computational module for each subsystem can be modeled as a discipline and numerical techniques are employed to handle their interactions while computing the system-level gradients that are used to find optimal design parameters with respect to some objective value \citep{simpson2011multidisciplinary, allison2014special, martins2021engineering}. Due to the complexities and computational time involved in such methods, low-fidelity discipline models are preferred when performing system-level MDO \citep{gray2014automatic}. For example, when optimizing a wind turbine system that involves the computation of forces applied on the rotor, one could employ the blade element momentum theory \citep{mahmuddin2017rotor} instead of performing a high-fidelity computational fluid dynamics analysis over the whole wind turbine to obtain the forces.

In many applications, however, some particular subsystem may warrant employing a high-fidelity model to compute its behaviour. For example in vehicle dynamics, while a suspension system could be represented using a simple mass-spring-damper model, it will not capture the behavioural differences between different types of suspension designs, e.g., a double-wishbone design vs. a MacPherson design. Not being able to capture such details may lead to inaccurate estimation of the overall system behaviour and subsequently sub-optimal design solutions.

In order to address this challenge, we propose a mixed-fidelity MDO approach, using the vehicle dynamics optimization problem as our case study. We employ low-fidelity vehicle dynamics equations that are typically used to model the overall behaviour of the vehicle, while incorporating suspension-wheel models using a high-fidelity multi-body dynamics solver. We demonstrate our approach with an experiment that compares the simulation results between the low-fidelity approach versus our mixed-fidelity approach, and the optimization results obtained by employing both approaches in a sequential manner.

To solve the optimization problem efficiently, we leverage the gradient-based techniques developed by the MDO community, namely MAUD (modular analysis and unified derivatives) \citep{hwang2018computational}, which allows us to compute the total derivatives of an objective function using the sensitivities obtained from the mixed-fidelity models. Also, we employ a collocation method \citep{herman1996direct}, which serves as the common time integration scheme for both low-fidelity and high-fidelity models, while formulating both the time integration problem and the design optimization problem as a single monolithic problem that can be solved efficiently \citep{falck2021dymos}. 

\section{PROBLEM: OPTIMAL SUSPENSION DESIGN FOR VEHICLE DYNAMICS}

\begin{figure}[t]
\centering
\includegraphics[width=12cm]{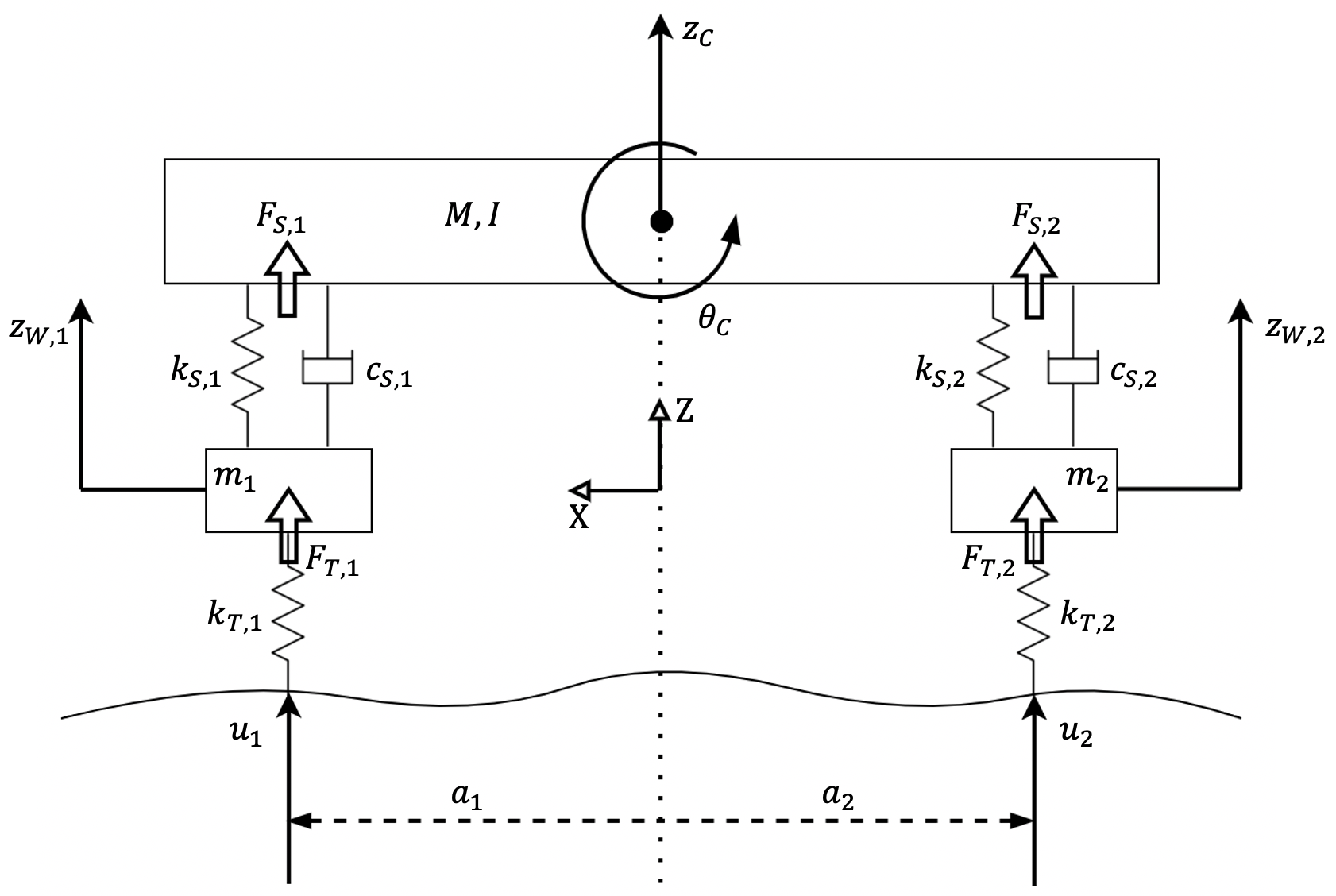}
\caption{Half-car model used for modeling vertical vehicle dynamics.}
\label{car}
\end{figure}

Optimal suspension design for vehicle dynamics can be defined as follows
\begin{equation}
\begin{aligned}
& \underset{x}{\text{min}}
& & f(u(x)) \\
& \text{s.t.}
& & G(u(x), x) = 0 \\
& & & L_i \leq x_i \leq U_i
& & & &i=1,2,...,|x_i|
\label{eq:opt}
\end{aligned}
\end{equation}
where $u$ is the vector of state variables defined for a particular car model. For the half-car model (Figure \ref{car}) used for the current work focusing on vertical dynamics, they are 
\begin{equation}
u = [z_C, \theta_C, z_{W,1}, z_{W,2}]
\end{equation}
which include the vertical displacement of the chassis $z_C$, the pitch angle of the chassis $\theta_C$, and the vertical displacements of the front and rear wheels, $z_{W,1}$ and $z_{W,2}$, respectively.

Next, $x$ is a set of design variables considered for the current work, defined as
\begin{equation}
x = [k_{S,1}, k_{S,2}, c_{S,1}, c_{S,2}]
\end{equation}
where $k_{S,1}$ and $k_{S,2}$ are the spring constants of the front and rear suspension systems, and $c_{S,1}$ and $c_{S,2}$ are the damping coefficients of the same systems, with imposed bounds $L_j$ and $U_j$.

The objective function $f$, can be defined as follows \citep{goga2012optimization}
\begin{equation}
f = \omega_1 \int (\ddot{z}_C)^2 dt + \omega_2 \int (\ddot{\theta}_C)^2 dt + \omega_3 \int (z_{W,1})^2 dt + \omega_4 \int (z_{W,2})^2 dt.
\label{obj}
\end{equation}
where the first two integrals capture the vertical and pitch angular accelerations of the chassis over time, which correspond to the passenger's discomfort. The last two integrals capture the vertical displacements of each wheel, which correspond to the contact between the wheels and the road that is critical for driving performance. This is a weighted sum of four objective terms, hence $\omega_i$ are the weights assigned for each term.

Note that the objective function is dependent on $u$ and their time derivatives, which must satisfy the set of governing equations $G(u(x), x)$. These equations essentially model the behaviours of the chassis, suspensions, and wheels, the exact forms of which are presented in the following section for different fidelity approaches.

\section{METHODS}

\subsection{Low-fidelity approach using half-car dynamics}

For the low-fidelity approach, we use the following set of ordinary differential equations (ODE) to model half-car dynamics \citep{rill2020road}. 
\begin{equation}
\label{motion1}
M \ddot{z}_C = F_{S,1} + F_{S,2},
\end{equation}
\begin{equation}
\label{motion2}
I \ddot{\theta}_C = -a_1 F_{S,1} - a_2 F_{S,2},
\end{equation}
\begin{equation}
\label{motion3}
m_1 \ddot{z}_{W,1} = -F_{S,1} + F_{T,1},
\end{equation}
\begin{equation}
\label{motion4}
m_2 \ddot{z}_{W,2} = -F_{S,2} + F_{T,2}.
\end{equation}
where $M$ and $I$ are the mass and the moment of inertia of the chassis, $a_1$ and $a_2$ are the front and rear axle distances from the center of mass of the chassis, and $m_1$ and $m_2$ are the mass of the front and rear wheels. We assume the suspension system is in a pre-loaded state so ignore gravity. 

The suspension forces $F_{S,i}$ and tire forces $F_{T,i}$ for each wheel $i$ are defined as follows
\begin{equation}
\label{susp}
F_{S,i} = k_{S,i} (z_{W,i} + a_i \theta_C - z_C) + c_{S,i} (\dot{z}_{W,i} + a_i \dot{\theta}_C - \dot{z}_C),
\end{equation}
\begin{equation}
\label{tire}
F_{T,i} = k_{T,i} (\delta_i - z_{W,i}).
\end{equation}
where $\delta_i$ is the road displacements applied at each wheel.

The advantages of this low-fidelity model are its small number of degrees-of-freedom (DOF) and that computing the analytical derivatives required for gradient-based optimization is straightforward. However, it does not capture the intricate details of a suspension system. 

\subsection{High-fidelity approach using multi-body dynamics}

An alternative approach to solving the vehicle dynamics problem is to use a multi-body dynamics (MBD) solver to model and analyze the whole vehicle. A multi-body dynamical system can be represented using the following system of equations \citep{shabana2020dynamics}

\begin{equation}
\begin{bmatrix}
\mathbf{M} & \mathbf{C}_q^T \\
\mathbf{C}_q & \mathbf{0}
\end{bmatrix}
\begin{bmatrix}
\ddot{\mathbf{q}} \\
\boldsymbol{\lambda}
\end{bmatrix}
=
\begin{bmatrix}
\mathbf{F}_e + \mathbf{F}_v \\
\mathbf{F}_c
\end{bmatrix}
\end{equation}
where $\mathbf{M}$ is the mass matrix of the system, $\mathbf{q}$ are the state variables or DOF of the system, $\mathbf{F}_e$ are the external forces, $\mathbf{F}_v$ are the velocity dependent forces for example due to the damping, and $\mathbf{F}_c$ are the constraint forces associated with the constraint equations between the bodies in the system. $\mathbf{C_q}$ is the constraint Jacobian matrix derived from the constraint equations and $\boldsymbol{\lambda}$ are their Lagrange multipliers applied to enforce those constraints. 

\subsection{Proposed approach: Mixed-fidelity}

\begin{figure}[t]
\centering
\includegraphics[width=10cm]{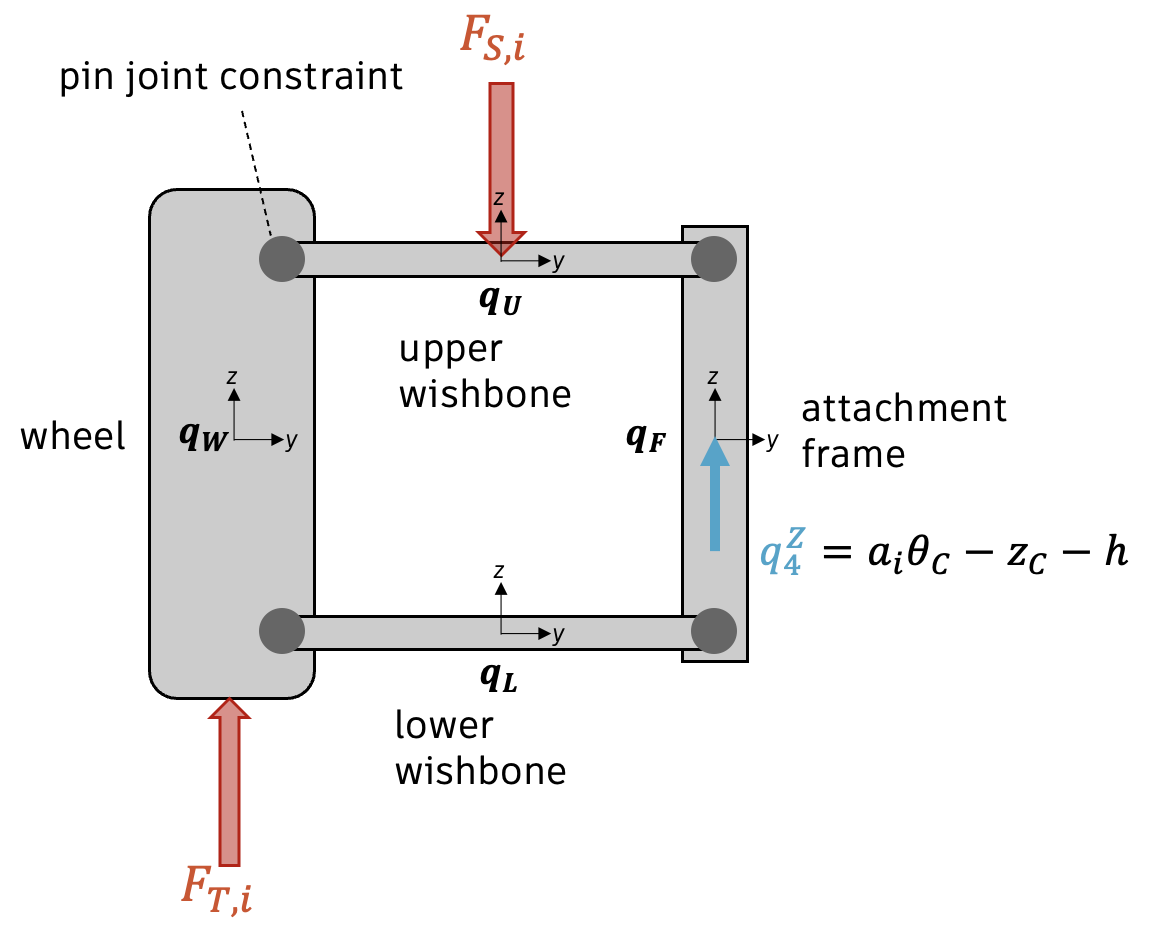}
\caption{Multi-body dynamics model used for the suspension-wheel assembly.}
\label{mbd}
\end{figure}

For the mixed-fidelity approach we propose, the suspension-wheel assembly is represented with a MBD model as shown in Figure \ref{mbd}, resembling a double-wishbone design. Hence, $\mathbf{q} = [\mathbf{q_W}, \mathbf{q_U}, \mathbf{q_L}, \mathbf{q_F}]$ and $\mathbf{F}_e = [\mathbf{F}_{S,i}, \mathbf{F}_{T,i}]$. The other two forces can be computed using $\mathbf{q}$ and $\mathbf{\dot{q}}$ \citep{shabana2020dynamics}. $\mathbf{M}$ can be constructed with the mass and the moments of inertia for each body in the system. Finally, we assume pin joint constraints for all the joints in the system and compute $\mathbf{C_q}$ based on $\mathbf{q}$ \citep{shabana2020dynamics}. Therefore, assuming $\mathbf{M}$ is fixed for the system, given $\mathbf{q}$, $\mathbf{\dot{q}}$, and $\mathbf{F}_e$ at a particular point in time, we can initialize and solve the above system of equations to obtain $\mathbf{\ddot{q}}$. Any capable MBD solver can be used here, and we use \citep{ebrahimi2019design} for this work.

For the mixed-fidelity approach, the system of MBD equations above replaces Eqs \eqref{motion3} and \eqref{motion4}, while the force equations \eqref{susp} and \eqref{tire} are replaced with the following equations, as depicted in Figure \ref{mbd}.
\begin{equation}
\label{susp2}
F_{S,i} = k_{S,i} (q^{z}_{U, i} + a_i \theta_C - z_C) + c_{S,i} (\dot{q}^{z}_{U, i}+ a_i \dot{\theta}_C - \dot{z}_C),
\end{equation}
\begin{equation}
\label{tire2}
F_{T,i} = k_{T,i} (\delta_i - q^{z}_{W,i}).
\end{equation}

\subsection{Collocation method}

All motion equations used above for vehicle dynamics are ODE's. Solving these equations over a time interval requires applying a time integration scheme. The common approach is to use an implicit time-marching method such as the Runge-Kutta method \citep{haug1999implicit} to solve for the states \emph{forward} in time. For design optimization, one would also need to solve for the adjoints \emph{backward} in time, which are necessary for computing the total sensitivities \citep{ebrahimi2019design}. However, the following challenges arise when applying this approach to our application.

First, note that a MBD solver employed to model and analyze the suspension-wheel assembly will have a particular time integration scheme built in. On the other hand, we also need a time integration scheme for Eqs \eqref{motion1} and \eqref{motion2}, which must be compatible with the one employed for the MBD solver. If there is no knowledge of the time integration method used in the MBD solver, this challenge is difficult to resolve.

Next, a MBD solver may only perform the forward simulation and not provide the adjoints (based on the backward simulation) that are required to compute the gradients for design optimization. This is true for almost all the commercially available MBD solvers.

To overcome these challenges, we use a collocation method. The collocation method \citep{garg2011direct} dicretizes the time domain into multiple segments and imposes defect constraints on each time segment defined between its discretization nodes. Then, a polynomial is approximated between the discretization nodes using their state values (guessed initially), and a defect constraint is defined as the difference in the slopes of the polynomial and the actual rates given by the ODE's at \emph{collocation nodes} within the time segment. The problem then becomes finding the state values at discretization nodes across the whole time interval that drive all the defect constraints to zero. This essentially turns a time integration problem into a nonlinear, constrained optimization problem, which can be efficiently solved using an optimizer tailored for such a problem \citep{falck2021dymos}.

In this approach, both the half-car dynamics equations and the MBD solver are used to evaluate the state rate values at the discretization and collocation nodes during the optimization. Hence, the collocation method serves as the common ``time integrator''. Also, we incorporate the design variables and objective function as part of the optimization problem, and therefore simultaneously solve for both state and design variables. To address the lack of adjoints provided by a MBD solver, we compute finite-difference gradients for the MBD model and combine them with the analytical gradients computed for vehicle dynamic equations, and these gradients form the total system Jacobian \citep{hwang2018computational} used during optimization.

\subsection{Final optimization problem}

The final optimization problem based on the mixed-fidelity model and the collocation method can be defined as follows
\begin{equation}
\begin{aligned}
& \underset{x, \Bar{u}}{\text{min}}
& & f = \sum^{N_T}_{i=1}{\omega_1 (\ddot{z}_{C})_i} + {\omega_2 (\ddot{\theta}_{C})_i} + {\omega_3 (q^{z}_{W,1})_i} + {\omega_4 (q^{z}_{W,2})_i}\\
& \text{s.t.}
& & \eta_j(\Bar{u}, \dot{\Bar{u}}) = 0 \\
& & & \zeta_j(\dot{\Bar{u}}, \ddot{\Bar{u}}(x, \Bar{u}, \dot{\Bar{u}})) = 0
& & & & j=1,2,...,(N_T-1) \\
& & & L_k \leq x_k \leq U_k
& & & & k=1,...,|x_j|
\label{final}
\end{aligned}
\end{equation}
where $\Bar{u}$ is the discretized set of state variables over time. The objective function is a discretized form of Eq \eqref{obj}. $\eta_j$ and $\zeta_j$ are the defect constraints involving the state variables, their first time derivatives, and their second time derivatives, the last of which are evaluated using the vehicle dynamics models. 

\subsection{Implementation details}

We use Dymos \citep{falck2021dymos} to implement our vehicle dynamics models and leverage its collocation method and optimization library (wrapped around IPOPT, Interior Point OPTimizer \citep{pirnay2012optimal}) to solve the optimization problem posed as Eq \eqref{final}. Dymos is an open-source Python library that enables simulation or optimization involving multidisciplinary time-dependent systems.

Each of the vehicle dynamics equations \eqref{motion1}, \eqref{motion2}, \eqref{motion3}, \eqref{motion4} as well as the two versions of the force equations \eqref{susp}, \eqref{tire} and \eqref{susp2}, \eqref{tire2} is modeled as a discipline \emph{component} in Dymos. Each component computes either the acceleration terms or the force terms as outputs given the other variables as inputs and also the partial derivatives of the outputs with respect to all the inputs. For the MBD solver, it is also wrapped as a Dymos component and outputs the acceleration terms given the other variables as inputs. However, because the solver does not provide any gradients, we used the finite-differencing feature available in Dymos to compute the partial derivatives. Dymos then uses MAUD to assemble all these partial derivatives to compute the system Jacobian providing the total derivatives of both the objective function and the defect constraint functions with respect to the design and state variables to the optimizer.  

\section{EXPERIMENT}

\subsection{Two-step solving approach}

To demonstrate the value of the mixed-fidelity approach, we devised the following experiment. The optimization problem defined in Eq. \eqref{final} is first solved using the low-fidelity approach to prioritize obtaining solutions quickly. The design variables are initialized with the mid point values between their bounds (to be presented). We then use the optimal values obtained as the initial guess for the mixed-fidelity approach, and attempt to find a better solution with a lower objective value. This two-step method resembles an engineering design process when an engineer goes from abstract to more concrete design and analysis.

\subsection{Optimization settings}

The parameter values used for the vehicle dynamics equations and force equations are listed in Table \ref{params}. Also, the parameter values used for the MBD model are listed in Table \ref{mbdparams}. Lastly, Table \ref{var} lists the lower and upper bounds used for the design variables.

\begin{figure}[h!]
\centering
\includegraphics[width=10.5cm]{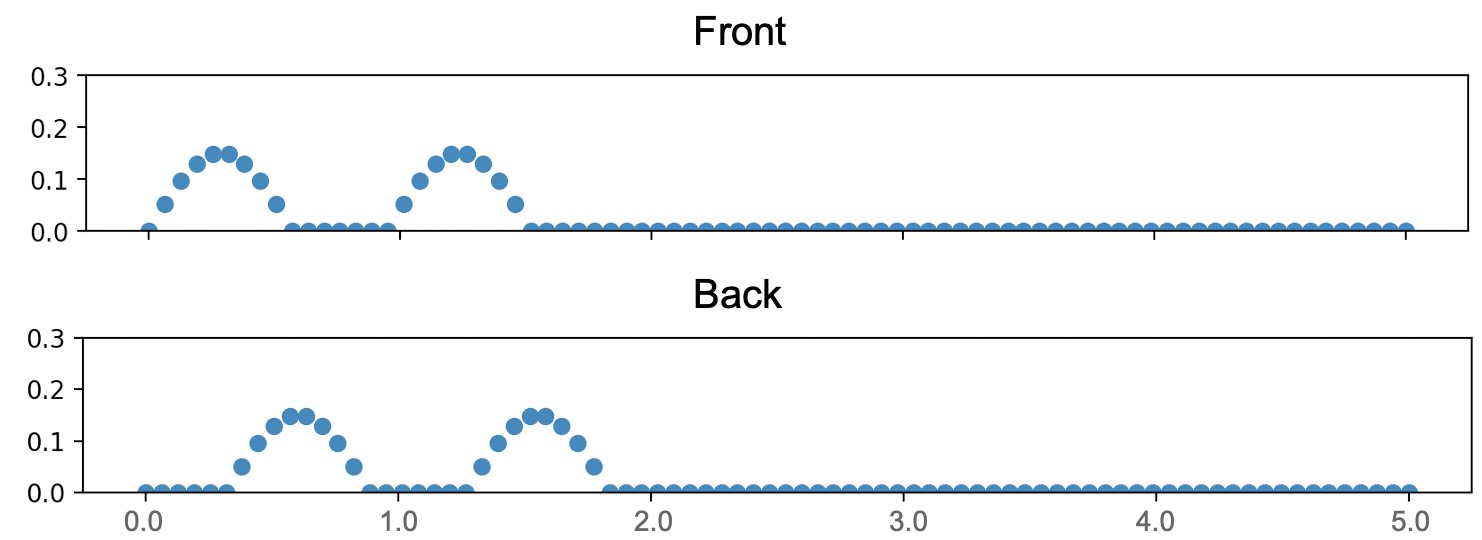}
\caption{Road displacements applied to each wheel for the experiment.}
\label{road}
\end{figure}

We apply road displacements at the front and back wheels that resemble a car going over two consecutive bumps, as shown in Figure \ref{road}. The time duration used for the problem was 5.0$s$. For the objective function in \eqref{final}, we normalize each term with scaling factors $0.1, 0.01, 1.0, 1.0$, respectively and apply equal weights $\omega_1=\omega_2=\omega_3=\omega_4=0.25$.

\begin{table}[h!]
\caption{\label{params} Parameter values for the vehicle dynamics and force equations used for the experiment.}
\centering
\begin{tabular}{c c|c c|c c}
$a_1$ & 1.271 m & $a_2$ & -1.713 m & $M$ & 1414 kg \\
$m_i$ & 120 kg & $k_{T,i}$ & 200000 N/m & $I$ & 2782 kg$\cdot$m$^2$ \\
\end{tabular}
\end{table}

\begin{table}[h!]
\caption{\label{mbdparams} Parameter values for the MBD model used for the experiment. The subscripts $W, F, U, L$ corresponds to those bodies with the same subscripts used in Figure \ref{mbd}. $I_x, I_y, I_z$ are the moments of inertia for each body. $p_{CG}$ are the positions of the center of gravity for each body. $p_{W:U}, p_{W:L}, p_{U:F},$ and $p_{U:L}$ are the joint locations between the bodies denoted with the subscripts. $p_{F_{S,i}}$ and $p_{F_{T,i}}$ are the locations where the corresponding forces are applied. Refer to Figure \ref{mbd} for the visualization of these locations. For all the position vectors, we ignore indicating the y-axis values as they are all zero.}
\centering
\begin{tabular}{c c|c c|c c|c c}
$m_{W,i}$ & 80 kg & $I_{x,W,i}$ &  4.14 kg$\cdot$m$^2$ & $I_{y,W,i}$ &  7.5 kg$\cdot$m$^2$ & $I_{z,W,i}$ & 4.14 kg$\cdot$m$^2$ \\
$m_{F,i}$ & 10 kg & $I_{x,F,i}$ & 0.517 kg$\cdot$m$^2$ & $I_{y,F,i}$ & 0.938 kg$\cdot$m$^2$ & $I_{z,F,i}$ & 0.517 kg$\cdot$m$^2$ \\
$m_{U,i}$ & 20 kg & $I_{x,U,i}$ & 0 kg$\cdot$m$^2$ & $I_{y,U,i}$ & 0.15 kg$\cdot$m$^2$ & $I_{z,U,i}$ & 0.15 kg$\cdot$m$^2$ \\
$m_{L,i}$ & 20 kg & $I_{x,L,i}$ & 0 kg$\cdot$m$^2$ & $I_{y,L,i}$ & 0.15 kg$\cdot$m$^2$ & $I_{z,L,i}$ & 0.15 kg$\cdot$m$^2$ \\
$p_{\text{CG},W,i}$ & [0.54, 0]m & $p_{\text{CG},F,i}$ & [0, 0]m & $p_{\text{CG},U,i}$ & [0.27, 0.2]m & $p_{\text{CG},L,i}$ & [0.27, -0.2]m\\
$p_{W:U}$ & [0.45, 0.2]m & $p_{W:L}$ & [0.45, -0.2]m & $p_{U:F}$ & [0.12, 0.2]m & $p_{U:L}$ & [0.12, -0.2]m\\
$p_{F_{S,i}}$ & [0.27, 0.2]m & $p_{F_{T,i}}$ & [0.54, -0.38]m & & & & \\
\end{tabular}
\end{table}

\begin{table}[h!]
\caption{\label{var} Design variable bounds used for the experiment.}
\centering
\begin{tabular}{c c|c c|}
$k_{S,1}$ & [10000, 50000] N/m &
$k_{S,2}$ & [10000, 50000] N/m \\
$c_{S,1}$ & [1000, 5000] N$\cdot$s/m& 
$c_{S,2}$ & [1000, 5000] N$\cdot$s/m \\
\end{tabular}
\end{table}

For the collocation method, we used the Radau pseudospectral option available in Dymos with the number of segments $=16$ and the polynomial order $=5$. We verified that this resolution was sufficient to produce the same trajectory results as a time-marching method.

For both steps of the two-step procedure, optimization is ran until it is stopped based on the default convergence criteria implemented in IPOPT.

\subsection{RESULTS}

\begin{figure}[b!]
\centering
\includegraphics[width=16cm]{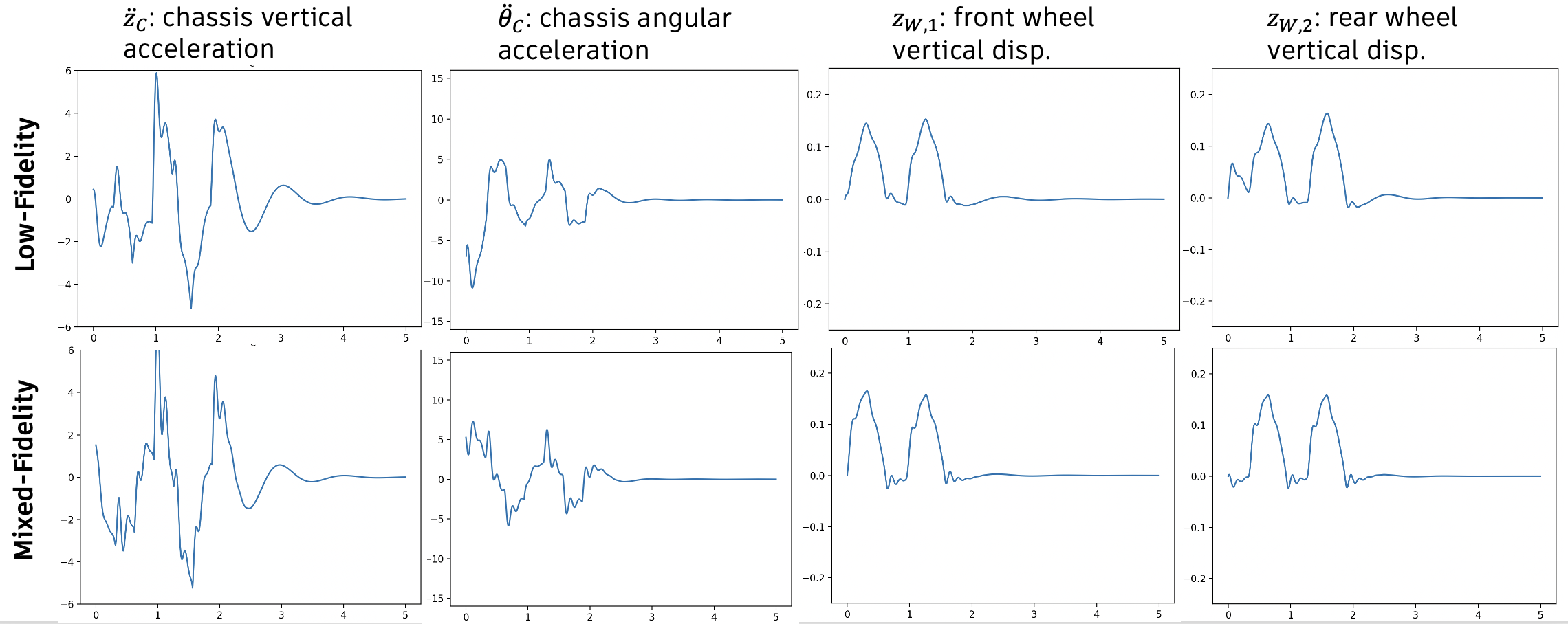}
\caption{Comparison of simulation results between the low-fidelity and mixed-fidelity approaches.}
\label{sim}
\end{figure}

We compare the simulation results obtained with the low-fidelity and the mixed-fidelity approaches by plotting the state variables of interest over time, as shown in Figure \ref{sim}. We can see that the mixed-fidelity model captures more intricate fluctuations of the state variables, especially for $\ddot{z}_C$ and $\ddot{\theta}_C$.

\begin{figure}[h!]
\centering
\includegraphics[width=16cm]{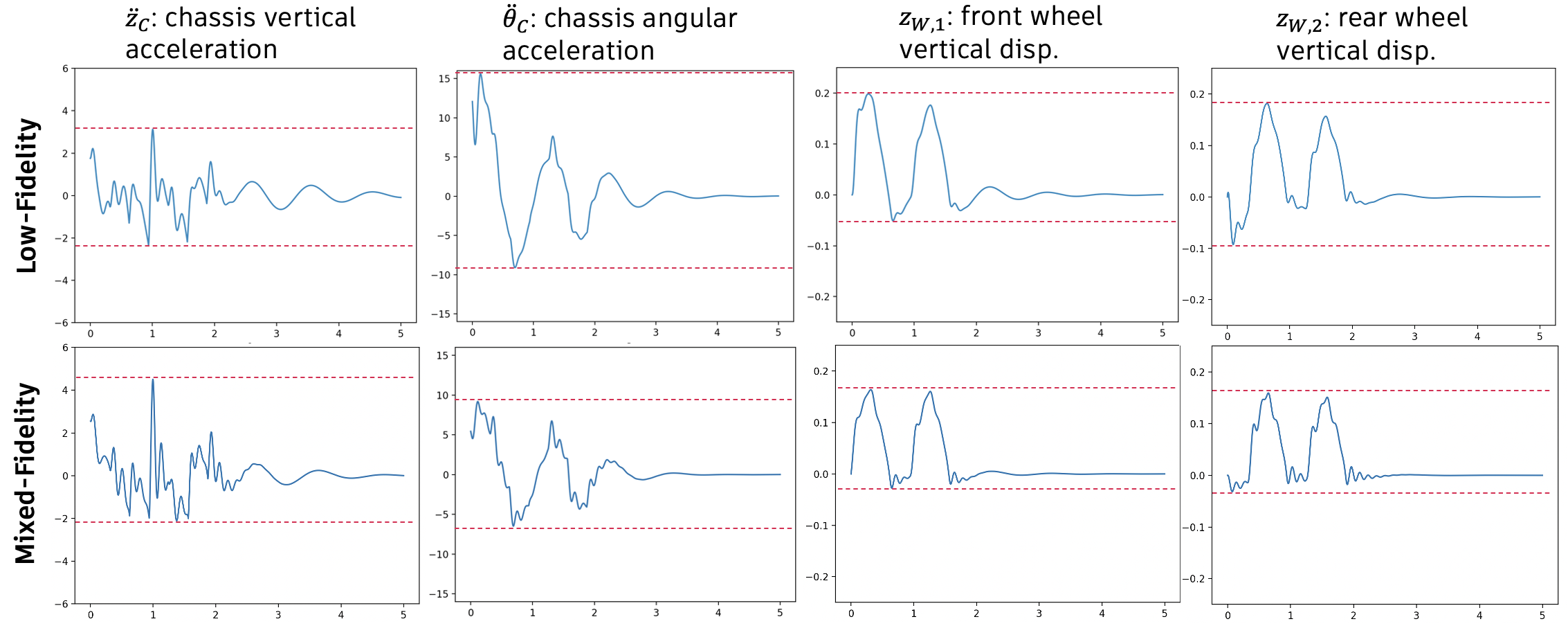}
\caption{Comparison of simulation results for the optimal solutions found with the low-fidelity and mixed-fidelity approaches.}
\label{opt}
\end{figure}

Next, we compare the simulation results for the optimal solutions found with each approach, as shown in Figure \ref{opt}. It can be seen that a significant improvement is made in minimizing the $\ddot{\theta}_C$ over time with the mixed-fidelity approach. 

Finally, Table \ref{res} shows the optimal values of the design variables and the objectives found for the low-fidelity and mixed-fidelity approaches, compared to the initial guess baseline. It is shown that optimizing with the low-fidelity approach first brings down the objective value from 0.909 to 0.384, while the subsequent mixed-fidelity approach further brings down the objective value to 0.322, demonstrating the benefit of the latter in improving the optimal solution found. 

\begin{table}[h!]
\caption{\label{res} Optimal values of the design variables and objective found for each approach, compared to the initial guess values.}
\centering
\begin{tabular}{c|cccc|c}
 & $k_{S,1} [N/m]$ & $k_{S,2}$ [N/m] & $c_{S,1}$ [N$\cdot$s/m] & $c_{S,2}$ [N$\cdot$s/m] & $f$ \\
 \hline
Initial guess & 30000 & 30000 & 3000 & 3000 & 0.909 \\
Low-fidelity & 42400 & 20700 & 2820 & 2010 & 0.384 \\
Mixed-fidelity & 37500 & 16400 & 2940 & 1890 & \textbf{0.322} \\
\end{tabular}
\end{table}

\section{SUMMARY AND CONCLUSIONS}

The current work demonstrates the application of a mixed-fidelity model for design optimization involving vehicle dynamics. In particular, we combine low-fidelity half-car dynamics model and a high-fidelity MBD solver to optimize suspension parameters for comfort and driving performance focusing on the vertical dynamics. We used the collocation method to solve for the state variables of both the low- and high-fidelity models and simultaneously include design optimization as a monolithic problem. The total gradients are computed out of the partial derivatives from each model based on MAUD, which are required for employing a gradient-optimization algorithm, namely IPOPT. We implemented our approach on Dymos, an open-source library that readily provides all these optimization techniques.

We highlight the value of the mixed-fidelity approach with an experiment featuring a two-step problem solving procedure. The low-fidelity approach is first used to find the initial optimal solution and the mixed-fidelity approach is subsequently used to improve the solution. The simulation results obtained with the mixed-fidelity approach also show more granular trajectories than the low-fidelity approach. 

In summary, the current work demonstrates that a gradient-based, mixed-fidelity multidisciplinary design optimization approach provides an effective solution for tackling a complex vehicle dynamics design application.

\newpage






\bibliographystyle{plain}
\bibliography{ref.bib}

\end{document}